\documentclass[%
reprint,
superscriptaddress, longbibliography
 amsmath,amssymb,
 aps,
]{revtex4-2}

\usepackage{graphicx}
\usepackage{dcolumn}
\usepackage{bm}
\usepackage{graphicx}
\usepackage{color,soul}
\usepackage{amsmath,amssymb}
\usepackage{mathrsfs}
\usepackage{color}
\usepackage{afterpage}
\usepackage{pifont}
\usepackage{xcolor}
\usepackage[version=3]{mhchem}
\usepackage{natbib}
\usepackage{soul}
\usepackage[caption=false]{subfig}
\usepackage{xtab,afterpage}
\usepackage{lipsum}
\usepackage{array}
\usepackage{multirow}

\begin{document}

\preprint{APS/123-QED}

\title{Tuning spin currents in collinear antiferromagnets and altermagnets}

\author{Sajjan Sheoran}
\affiliation{Department of Physics and Astronomy, Howard University, Washington D.C., USA}
\author{Pratibha Dev}%
 \email{pdev@lps.umd.edu}
\affiliation{Department of Physics and Astronomy, Howard University, Washington D.C., USA}
\affiliation{Laboratory for Physical Sciences, College Park, Maryland 20740, United States}
\date{\today}

\begin{abstract}
Spin current generation through non-relativistic spin splittings, found in uncompensated magnets and $d$-wave altermagnets, is desirable for low-power spintronics.  Such spin currents, however, are symmetry forbidden in conventional collinear antiferromagnets and higher-order altermagnets. Using spin point group analysis, we demonstrate that finite spin currents can be induced in these materials via magnetoelectric, piezomagnetic, and piezomagnetoelectric-like couplings. We utilize electric fields, strain, and their combinations to  drive symmetry-lowering phase transitions into uncompensated magnetic or $d$-wave altermagnetic states, thereby enabling finite spin conductivity in a broader class of magnetic materials.  We further substantiate this framework using density functional theory and Boltzmann transport calculations on representative magnetic materials -- KVSe$_2$O, RuF$_4$, Cr$_2$O$_3$, FeS$_2$, and MnPSe$_3$ -- spanning these different cases. The charge-to-spin conversion ratio reaches up to almost 100\% via uncompensated magnetism and about 40\% via $d$-wave altermagnetism under realistic conditions, highlighting the effectiveness of this approach for efficient spin current generation.
\end{abstract}

\maketitle


\section{Introduction} 
The generation of pure spin currents is of fundamental interest in spintronics.  In most materials, spin currents are typically achieved via the spin-Hall effect driven by spin-orbit interaction (SOI)~\cite{vzutic2004spintronics, hirsch1999spin}. However, SOI, which gives rise to spin currents, also results in spin dephasing mechanisms, such as D'yakonov-Perel~\cite{dyakonov1972spin} and Elliott-Yafet processes~\cite{elliott1954theory, yafet1963solid}. Such dephasing mechanisms limit the coherence length of the spin currents, thereby constraining their practical applications. Recently, collinear compensated magnets~\cite{hayami2020bottom,yuan2021prediction,vsmejkal2022beyond,sheoran2025spontaneous} and noncollinear antiferromagnets~\cite{vzelezny2017spin, flebus2019interfacial, cheong2024altermagnetism} have been shown to generate highly spin-polarized currents through exchange fields. The charge-to-spin conversion efficiency and coherence length of such nonrelativistic spin currents can exceed those of SOI-driven mechanisms by orders of magnitude~\cite{gonzalez2021efficient, vakili2025spin, bose2022tilted}.

In the context of symmetry, collinear compensated magnets are classified as antiferromagnets (AFMs) or altermagnets (AMs), depending on the symmetries that connect opposite spin sublattices~\cite{vsmejkal2022beyond}. In AFMs, the opposite spin sublattices are related by spatial inversion ($P$) and translation ($\tau$), leading to spin-degeneracy and zero spin conductivity. In contrast, AMs have mirror-rotation symmetries connecting opposite spin sublattices, and are further categorized by their planar or bulk $d$-, $g$-, and $i$-wave orders. Spin current generation has been theoretically predicted in $d$-wave AMs, such as RuO$_2$~\cite{gonzalez2021efficient}. However, the altermagnetic ground state of RuO$_2$ is still under debate~\cite{bose2022tilted}. Unfortunately, spin conductivity remains forbidden in higher-order AMs ($g$- and $i$-wave) due to additional spin-degenerate nodal surfaces. Nevertheless, one can use strain and electric fields to manipulate the altermagnetic order to allow for spin currents~\cite{belashchenko2025giant,karetta2025strain, vsmejkal2024altermagnetic}. For instance, recent works used uniaxial strain to induce $g$- to $d$-wave altermagnetic phase transitions in MnTe~\cite{belashchenko2025giant} and CrSb~\cite{karetta2025strain}. Other researchers showed that an electric field applied to the antiferromagnetic MnPSe$_3$ can induce a transition to either a $d$-wave~\cite{wang2024electric} or an $i$-wave AM phase~\cite{mazin2023induced}. While most such studies have focused on altermagnetic spin splittings, a general framework for generating nonrelativistic spin currents via magnetic phase transitions has remained unexplored.

In this article, we provide a general approach based on spin group symmetry formalism to induce spin currents through magnetoelectric, piezomagnetic, and piezomagnetoelectric-like coupling. We show that electric ($\mathcal{E}_i$) and strain ($\eta_{ij}$) fields and their combination ($\mathcal{E}_i\eta_{jk}$) can be utilized to induce spin-currents in AFMs and higher-order AMs via reduction of their respective symmetries. By considering the lower effective spin point groups (SPGs) under external stimuli for all possible AFMs and higher-order AMs, we determine the correlation between spin current and $\mathcal{E}_i$, $\eta_{jk}$, and $\mathcal{E}_i\eta_{jk}$. We compute the spin-polarized conductivity for several representative examples of collinear compensated magnets using DFT and Boltzmann transport simulations. We also determine the spin conductivity for the reduced SPGs when the materials are subjected to an external electric field along an arbitrary direction.

\section{Computational details}
In the nonrelativistic case, conductivity can be described through the spin-up ($s_k$) and spin-down ($-s_k$) channels with spin polarization along the $k$ direction. Spin-polarized conductivity, $\sigma_{ij}^{s_k}$, is defined as $J_i^{s_k}=\sigma_{ij}^{s_k}\mathcal{E}_j$, where $\mathcal{E}_j$ is the electric field along the $j$-direction and $J_i^{s_k}$ is the current along the $i$-direction. Within the Boltzmann transport theory, the conductivity $\sigma_{ij}^{s_k}$ (contributed by $E_{n\textbf{k}}$) is expressed as~\cite{callaway2013quantum}
\begin{equation}
\sigma^{s_k}_{ij}(E_F) = -\frac{e^2 \tau}{8\pi^3 \hbar^2} \sum_n \int \frac{\partial E^{s_k}_{nk}}{\partial k_i} \frac{\partial E^{s_k}_{nk}}{\partial k_j} \frac{\partial f^0}{\partial E^{s_k}_{nk}} \, d^3k
\end{equation}
Here, $\tau$ is the relaxation time, and $f^0$ is the Fermi-Dirac distribution function. Within Eq. 1, the net charge and spin conductivity, respectively, are defined as
\begin{equation}
\begin{aligned}
	\sigma_{ij} &=  \sum_{k}\sigma^{s_k}_{ij}+\sigma^{-s_k}_{ij} \\
	\sigma^k_{ij} &=\frac{\hbar}{2e}(\sigma^{s_k}_{ij}-\sigma^{-s_k}_{ij}).
\end{aligned}
\end{equation}
The spin-polarized conductivity $\sigma^{s_k}_{ij}$ transforms under time-reversal ($\mathcal{T}$) as $\mathcal{T}\sigma^{s_k}_{ij}\rightarrow \sigma^{-s_k}_{ij}$. Accordingly, the charge and spin conductivity tensors are even and odd under time-reversal symmetry ($\mathcal{T}$), respectively.  We initialize the spin-quantization axis along the $z$-direction. Since the effect of SOI is expected to be negligible, it is excluded from the simulation~\cite{gonzalez2021efficient}. The absence of SOI means there is no spin canting, and consequently, $\sigma^{x/ y}_{ij}= 0$. This can also be seen from Equation 1: $\sigma^{s_k}_{ij}$ transforms as $s_kk_ik_j$. The spin-only symmetry $[C_z(\pi)||E]$ transforms $\sigma^{s_{x}/s_{y}}_{ij}$ to $\sigma^{-s_{x}/-s_{y}}_{ij}$, which implies that the net spin current, $\sigma^{x/y}_{ij}$, must vanish while the charge current can be nonzero. Therefore, under nonrelativistic collinear magnetism, the only allowed nonzero spin conductivity components are $\sigma^{z}_{ij}$. The shape of $\sigma^{z}_{ij}$ is further restricted by the nontrivial spin group symmetry. 

Density functional theory calculations were performed using the plane-wave ultrasoft pseudopotential~\cite{vanderbilt1990soft} approach implemented in Quantum ESPRESSO~\cite{giannozzi2009quantum}. The generalized gradient approximation with Hubbard correction (PBE+U) was used to treat exchange correlations self consistently~\cite{perdew1996generalized}. The Kubo formalism, combined with DFT, was used to determine the charge and spin conductivities. To do so, we constructed a tight-binding Hamiltonian in the atom-centered Wannier basis~\cite{marzari1997maximally} using the Wannier90 code~\cite{mostofi2008wannier90}.  In the single-particle energy eigenstate representation, the spin conductivity within the Kubo formalism is given by:
\begin{equation}
\begin{aligned}
\sigma^{k}_{ij} = &- \frac{e\hbar}{2\Gamma} V \int \frac{d^3\mathbf{k}}{(2\pi)^3} \sum_n 
\langle n\mathbf{k} | \hat{J}^k_i | n\mathbf{k} \rangle 
\langle n\mathbf{k} | \hat{v}_j | n\mathbf{k} \rangle \\
&\times \delta(E_{n\mathbf{k}} - E_F) 
\end{aligned}
\end{equation}
where the $\hat{J}^k_i=\frac{1}{2}\{\hat{v}_i, \hat{s}_k\}$ is the spin current operator. $\Gamma=\frac{\hbar}{2\tau}$ is a constant scattering rate, and is inversely proportional to the relaxation time, $\tau$. The replacement of $\hat{J}^k_i$ with $\hat{v}^k_i$ in Eq. 3 gives the charge conductivity. The conductivity tensors were obtained by solving Eq. 3 as a postprocessing step of DFT calculations on a dense $350^3$ $k$-mesh for 3D materials and $350^2$ $k$-mesh for 2D materials, using the Wannier-Linear-Response code~\cite{gonzalez2021efficient}. We made use of the Findsym~\cite{stokes2005findsym}, Bilbao Crystallographic Server~\cite{aroyo2006bilbao}, Findspingroup~\cite{chen2024enumeration}, and Tensor Symmetry~\cite{xiao2025tensorsymmetry} packages to obtain the nontrivial SPGs and the corresponding shape of $\sigma^{z}_{ij}$.

\begin{figure}
	\includegraphics[width=3.3in]{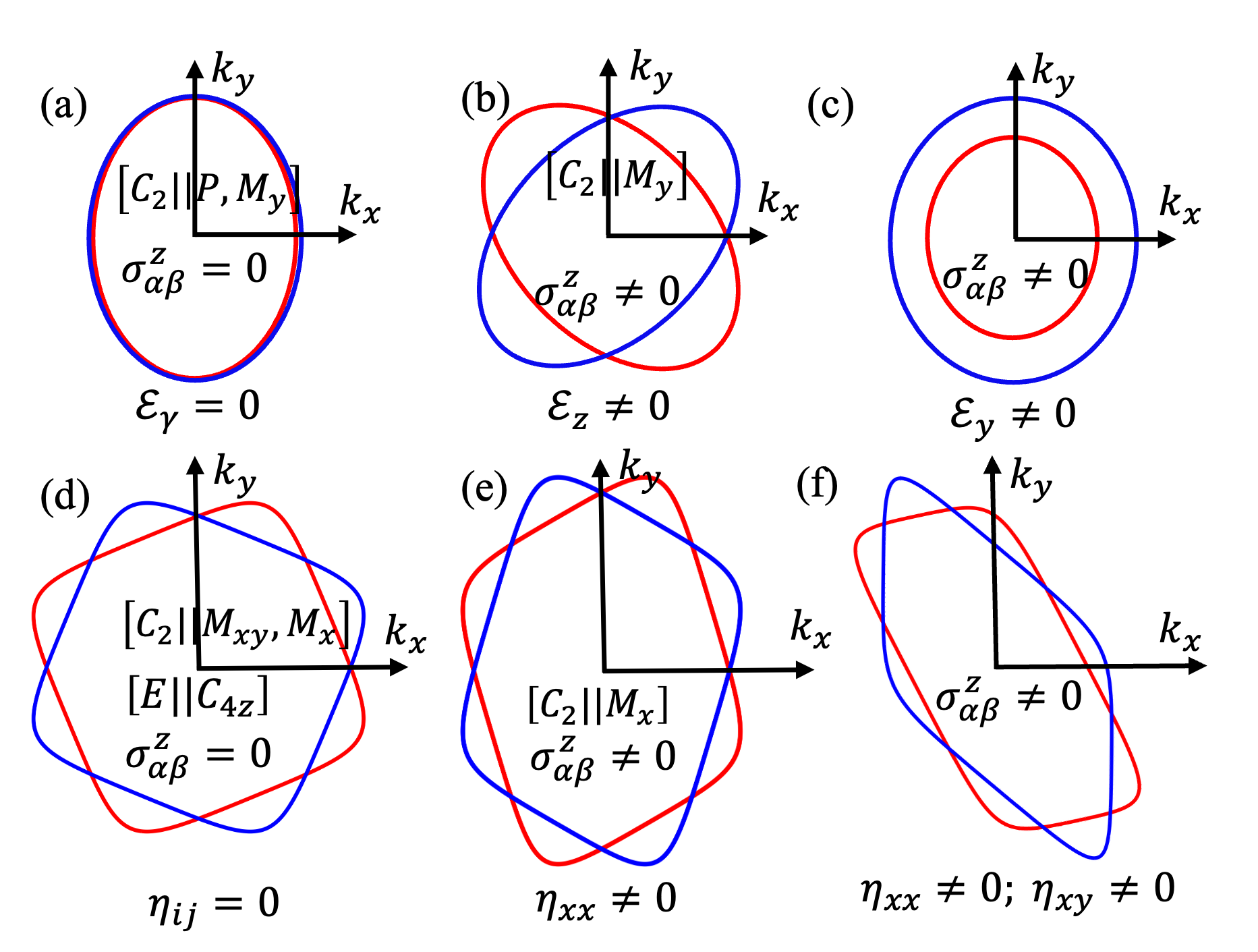}
	\caption{Schematic of electric field and strain-induced spin conductivity. (a) Spin-polarized Fermi surface of an antiferromagnet with $[C_2||P]$ symmetry. The red and blue isosurfaces correspond to the opposite spin directions. Fermi surfaces of an antiferromagnet under an applied electric field when: (b) the effective spin group of the material possesses a symmetry connecting opposite spin sublattices (e.g., $d$-wave altermagnet with $[C_2||M_y]$), and (c) no such symmetry exists (ferromagnetic or compensated ferrimagnetic). (d) Fermi surface of a higher-order planar $g$-wave state. Fermi surfaces of a higher-order $g$-wave altermagnet (d) without strain and under a strain field $\eta_{ij}$ when: (e) the effective spin group retains a symmetry connecting opposite spin sublattices (e.g., $d$-altermagnet with $[C_2||M_{xy}]$), and (f) no such symmetry remains. In the plots, Fermi surfaces are obtained from the minimal two-band models derived in Ref.~\cite{roig2024minimal}, with strain effects incorporated using $\textit{k}_i \rightarrow  (\delta_{ij}+\eta_{ij})\textit{k}_j$.}
    \label{p_s}
\end{figure}

\begin{table}
	\caption{Electric field-induced transition of antiferromagnetic SPG to ferromagnetic (or uncompensated magnetic) and altermagnetic SPG. The first column denotes the parent SPG, and the second, third, and fourth columns denote the effective SPG under the application of electric field $\mathcal{E}_x$, $\mathcal{E}_y$, and $\mathcal{E}_z$, respectively. The superscripts $1$ and $2$ preceding each symmetry operation indicate whether the spin sublattices are connected to the same or opposite spin sublattices. The SPG without any symmetry connecting opposite spin sublattices (SPGs with only superscript 1) are ferromagnetic. For the altermagnetic SPGs, the spin-orders ($d$, $g$, and $i$) are  also indicated in parentheses. }
	\centering
	\renewcommand{\arraystretch}{1.1}
	\begin{tabular}{ c|c c c}
		\hline \hline
		\textbf{SPGs} & $\mathcal{E}_x$ & $\mathcal{E}_y$ & $\mathcal{E}_z$ \\
		\hline 
		$^2{\overline{1}}$ & $^1 1$ & $^1 1$ & $^1 1$ \\
		$^22/^1m$ & $^1m$ & $^22$ ($d$) & $^1m$ \\
		$^12/^2m$ & $^2m$ ($d$) & $^12$ & $^2m$ ($d$) \\
		$^1 m ^2 m ^1m$ & $^22^2m^1m$ ($d$)  & $^1m^12^1m$ ($d$)  & $^1m^2m^22$ ($d$) \\
		$^1 m ^1 m ^2m$ & $^22^1m^2m$ ($d$)  & $^1m^22^2m$ ($d$)  & $^1m^1m^12$ ($d$) \\
		$^2 m ^2 m ^2m$  &  $^12^2m^2m$ ($d$) & $^2m^12^2m$ ($d$)  & $^2m^2m^12$ ($d$) \\
		$^14/^2m$ & $^2m$ ($d$)   & $^2m$ ($d$) & $^14$ \\
		$^24/^2m$ & $^2m$ ($d$)   & $^2m$ ($d$)  & $^24$ ($d$)  \\
		$^14/^2m ^1m ^1m$   & $^22^1m^2m$ ($d$)  & $^1m^22^2m$ ($d$)  & $^14^1m^1m$   \\
		$^24/^2m ^1m ^2m$ & $^22^1m^2m$ ($d$)  & $^1m^22^2m$ ($d$)  & $^24^1m^2m$ ($d$) \\
		$^14/^2m ^2m ^2m$  & $^12^2m^2m$ ($d$)  & $^2m^12^2m$ ($d$)  & $^14^2m^2m$  ($g$) \\
		$^2\bar{3}$ & $^11$ & $^11$ & $^13$ \\
		$^2\bar{3} ^11 ^1m$ & $^1m$ & $^22$ ($d$)& $^13^11^1m$ \\
		$^2 \bar{3} ^11 ^2m$ & $^2m$ ($d$) & $^12$ & $^13^11^2m$ ($g$)\\
		$^26/^1m$ & $^1m$  & $^1m$ & $^26$ ($g$) \\
		$^16/^2m$ & $^2m$ ($d$) & $^2m$ ($d$)  & $^16$ \\
		$^16/^2m^1m ^1m$ & $^22^1m^2m$ ($d$) & $^1m^22^2m$ ($d$) & $^16^1m^1m$  \\
		$^26/^1m^1m ^2m$ & $^22^2m^1m$ ($d$) & $^1m^12^1m$ & $^26^1m^2m$ ($g$) \\
		$^16/^2m^2m ^2m$& $^12^2m^2m$ ($d$) & $^2m^12^2m$ ($d$) & $^16^2m^2m$ ($i$) \\
		$^2m^2\overline{3}$ & $^12^2m^2m$ ($d$) & $^2m^12^2m$ ($d$) & $^2m^2m^12$ ($d$) \\
		$^2m^2\overline{3}{^1}m$& $^24^2m^1m$ ($d$)  & $^24^2m^1m$ ($d$) & $^24^2m^1m$ ($d$) \\
		$^2m^2\overline{3}{^2}m$ & $^14^2m^2m$ ($g$) & $^14^2m^2m$ ($g$) & $^14^2m^2m$ ($g$) \\
		\hline \hline
	\end{tabular}
\end{table}

\begin{table*}[t]
	\caption{Tuning altermagnetism by the application of strain. The first and second columns denote the altermagnetic pattern and the parent SPG, respectively. The latter six columns denote the effective SPG of the parent space group after the application of the strain field $\eta_{ij}$.}
	\centering
	\renewcommand{\arraystretch}{1.2}
	\begin{tabular}{c|c|c c c c c c }
		\hline \hline
		Altermagnetic Pattern	& \textbf{SPGs} & $\eta_{xx}$ & $\eta_{yy}$ & $\eta_{zz}$ & $\eta_{xy}$ & $\eta_{xz}$ & $\eta_{yz}$ \\\hline
		Planar $g$-wave	&	$^14/^1m^2m^2m$  & $^2m^2m^1m$ $(d)$ & $^2m^2m^1m$ $(d)$ & $^14/^1m^2m^2m$ $(g)$  & $^2m^2m^1m$ $(d)$ & $^22/^2m$ $(d)$ & $^22/^2m$ $(d)$ \\
		Planar $g$-wave	   & $^14^2m^2m$	& $^2m^2m^12$ $(d)$ & $^2m^2m^12$ $(d)$ & $^14^2m^2m$ $(g)$ & $^2m^2m^12$ $(d)$ & $^2m$ $(d)$ & $^2m$ $(d)$\\
		Planar $g$-wave	   & $^14^22^22$ & $^22^22^12$ $(d)$ & $^22^22^12$ $(d)$ & $^14^22^22$ $(g)$ & $^22^22^12$ $(d)$ & $^22$ $(d)$ & $^22$ $(d)$ \\
		Planar $g$-wave	   & $^1\overline{4} {^2}2^2m$ & $^22^22^12$ $(d)$ &  $^22^22^12$ $(d)$& $^1\overline{4} {^2}2^2m$ $(g)$ & $^2m^2m^12$ $(d)$& $^22$ $(d)$ & $^22$ $(d)$\\
		Bulk $g$-wave	 &  $^1\overline{3}{^2}m^11$& $^22/^2m$ $(d)$ & $^22/^2m$ $(d)$ &  $^1\overline{3}{^2}m^11$ $(g)$ & $^1\overline{1}$ & $^1\overline{1}$ &  $^22/^2m$ $(d)$\\
		Bulk $g$-wave	  &  $^13{^2}m^11$ & $^2m$ $(d)$ & $^2m$ $(d)$ & $^13{^2}m^11$ $(g)$ & $^11$ & $^11$ & $^2m$ $(d)$\\
		Bulk $g$-wave	  &  $ ^13^22^11$ & $^22$ $(d)$ & $^22$ $(d)$& $ ^13^22^11$  $(g)$ & $^11$ & $^11$ & $^22$ $(d)$ \\
		Bulk $g$-wave	  &  $^26/^2m$& $^22/^2m$ $(d)$ & $^22/^2m$ $(d)$ & $^26/^2m$ $(g)$ &  $^22/^2m$ $(d)$ & $^1\overline{1}$ & $^1\overline{1}$\\
		Bulk $g$-wave	  &  $^2\overline{6}$ & $^2m$ $(d)$ & $^2m$ $(d)$ & $^2\overline{6}$ $(g)$ & $^2m$ $(d)$ & $^11$ & $^11$ \\
		Bulk $g$-wave	  &  $^26$ & $^22$ $(d)$ & $^22$ $(d)$ & $^26$ $(g)$ & $^22$ $(d)$ & $^11$& $^11$ \\
		Bulk $g$-wave		  &  $^26/^2m^2m^1m$ & $^2m^1m^2m$ $(d)$ & $^2m^1m^2m$ $(d)$ & $^26/^2m^2m^1m$ $(g)$ & $^22/^2m$ $(d)$ & $^12/^1m$ & $^22/^2m$ $(d)$\\
		Bulk $g$-wave		 &   $^26^2m^1m$ & $^2m^1m^22$ $(d)$ & $^2m^1m^22$ $(d)$ & $^26^2m^1m$ $(g)$ & $^22$ $(d)$ & $^1m$ & $^2m$ $(d)$ \\
		Bulk $g$-wave		 &   $^26^22^12$& $^22^12^22$ $(d)$ & $^22^12^22 $ $(d)$ & $^26^22^12$ $(g)$ & $^22$ $(d)$ & $^12$ & $^22$ $(d)$ \\
		Bulk $g$-wave		 &   $^2\overline{6}{^2}m^12$ & $^2m^12^2m$ $(d)$ & $^2m^12^2m$ $(d)$ & $^2\overline{6}{^2}m^12$ $(g)$ & $^2m$ $(d)$ & $^12$ & $^2m$ $(d)$ \\
		Bulk $g$-wave		  &  $^2\overline{6}{^2}2^1m$& $^22^1m^2m$ $(d)$ & $^22^1m^2m$ $(d)$ & $^2\overline{6}{^2}2^1m (g)$ & $^2m$  $(d)$ & $^1m$ & $^22$ $(d)$ \\
		Planar $i$-wave		  &  $^16/^1m^2m^2m$ & $^2m^2m^1m$ $(d)$ & $^2m^2m^1m$ $(d)$ & $^16/^1m^2m^2m$ $(d)$ & $^12/^1m$ & $^22/^2m$ $(d)$ & $^22/^2m$ $(d)$ \\
		Planar $i$-wave		 &   $^16^2m^2m$ & $^2m^2m^12$ $(d)$ & $^2m^2m^12$ $(d)$ & $^16^2m^2m$ $(i)$ & $^12$ & $^2m$ $(d)$ & $^2m$ $(d)$\\
		Planar $i$-wave		 &   $^16^22^22$& $^22^22^12$ $(d)$ & $^22^22^12$ $(d)$ & $^16^22^22$ $(i)$ & $^12$ & $^22$ $(d)$ & $^22$ $(d)$\\
		Planar $i$-wave		 &   $^1\overline{6}{^2}m^22$& $^2m^22^1m$ $(d)$ & $^2m^22^1m$ $(d)$ & $^1\overline{6}{^2}m^22$ $(i)$ & $^1m$ & $^22$ $(d)$ & $^2m$ $(d)$ \\
		Bulk $i$-wave & $^1m^1\overline{3}{^2}m$ & $^24/^1m^1m^2m$ $(d)$ & $^24/^1m^1m^2m$ $(d)$ & $^24/^1m^1m^2m$ $(d)$ & $^2m^2m^1m$ $(d)$ & $^2m^2m^1m$ $(d)$ & $^2m^2m^1m$ $(d)$ \\
		Bulk $i$-wave & $^2\overline{4} {^1}3^2m$ & $^2\overline{4}{^1}2^2m$ $(d)$ & $^2\overline{4}{^1}2^2m$ $(d)$ & $^2\overline{4}{^1}2^2m$ $(d)$ & $^2m^2m^12$ $(d)$ & $^2m^2m^12$ $(d)$ & $^2m^2m^12$ $(d)$ \\
		Bulk $i$-wave & $^24{^1}3^22$ & $^24^12^22$ $(d)$ & $^24^12^22$ $(d)$ & $^24^12^22$ $(d)$ & $^22^22^12$ $(d)$ & $^22^22^12$ $(d)$ & $^22^22^12$ $(d)$ \\
		
		\hline \hline
	\end{tabular}
\end{table*}

\section{Symmetry Analysis}
Collinear magnets are more accurately described using spin group symmetry operations of the form $[R_1|| R_2]$, where $R_1$ and $R_2$ act on spin and real space, respectively~\cite{litvin1974spin}. These systems exhibit symmetries that fall into 32 SPGs associated with FM order, 22 SPGs corresponding to conventional AFM order, and 36 SPGs belong to AM order~\cite{chen2024enumeration}. The spin group operations transform $\sigma^{s_k}$ as 
\begin{equation}
	\mathbf{D}(R_2) \sigma^{\mathbf{D}(R_1)s_k} \mathbf{D}(R_2)^{\dagger} = \sigma^{s_k}
\end{equation}
The spin conductivity component, $\sigma^{z}_{ij}$,  will be allowed if the SPG does not contain any symmetry operation enforcing $\sigma^{s_z}_{ij}=\sigma^{-s_z}_{ij}$. Since the $[C_2||P]$ symmetry in AFMs leads to $\sigma^{s_k}=\sigma^{-s_k}$, the spin current $\sigma^{k}$ is forbidden for all AFM SPGs, as schematically shown in Fig.~\ref{p_s}(a) for SPGs containing $[C_2||P]$ and $[C_2||M_y]$. The $[C_2||P]$ symmetry can be broken by applying an electric field ($\mathcal{E}_i$) since $[C_2||P]\mathcal{E}_i\rightarrow -\mathcal{E}_i$.  The applied electric field ($\mathcal{E}_i$) also breaks the $[E/C_2||M_i]$ and $[E/C_2||C_{nj} (j\ne i)]$ symmetries.  Overall, the spin conductivity component, $\sigma^{z}_{ij}$, becomes allowed if the electric field breaks all spin group symmetries that enforce $\sigma^{s_z}_{ij}=\sigma^{-s_z}_{ij}$.  Table I lists the electric field-induced transitions of antiferromagnetic SPGs to either FM or AM SPGs. If the reduced effective SPG is an FM or a $d$-wave AM, a net spin current becomes allowed. In such a case, spin-polarized conductivity can be considered to be bilinear in the electric field. This coupling between spin current and electric field is reminiscent of magnetoelectric coupling. For instance, AFM SPG $^12/^2m$ (two-fold axis along $y$-direction) reduces to the effective AM SPG of $^2m$ under an electric field  either applied along $\mathcal{E}_x$ or $\mathcal{E}_z$. This case corresponds to the spin-polarized Fermi surface are shown schematically in Fig.~\ref{p_s}(b). Also, AFM SPG $^12/^2m$ reduces to an effective FM SPG of $^12$ under electric field $\mathcal{E}_y$ [spin-polarized Fermi surfaces in Fig.~\ref{p_s}(c)]. Table I also lists the cases where the applied electric field reduces the AFM SPG to that of either a $g$- or an $i$-wave AM.  For example, $\mathcal{E}_z$ applied to a $^14/^2m ^2m ^2m$ or $^2$6/$^1$m AFM, and $\mathcal{E}_i$ ($i=x, y, z$) applied to a  $^2m^2\overline{3}{^2}m$ AFM results in $g$-wave AMs.  Also, the $^16/^2m^2m ^2m$ AFM reduces to an $i$-wave AM under the application of $E_z$. In such cases, a higher order (i.e., trilinear) electric field may be required to induce a finite spin conductivity~\cite{ezawa2025third}.


Unlike the electric field that breaks the $[C_2||P]$ symmetry of AFMs, the strain field ($\eta_{ij}$, with $i, j=x, y, z$), which transforms like $k_ik_j$, remains invariant under $[C_2||P]\eta_{ij}\rightarrow\eta_{ij}$. Therefore, $\eta_{ij}$ is not sufficient to induce spin currents in AFMs.  However, strain can be used to induce phase transitions in $g$- and $i$-wave AMs to obtain either an FM or a $d$-wave AM. For example, AM SPG $^14^2m^2m$ is a $g$-wave altermagnet [see Fig.~\ref{p_s}(d)]. The $[E||C_{4z}]$, $[E||M_{x}]$, and $[E||M_{xy}]$ symmetries would forbid a spin current in such a material. However, the application of the shear strain $\eta_{xy}$ breaks $[C_2||M_{x}]$, leading to effective SPG $^2m^2m^12$ with spin-polarized Fermi surface similar to a $d$-wave AM as shown in Fig.~\ref{p_s}(e). In addition, a simultaneous application of $\eta_{xx}$ or $\eta_{yy}$ along with $\eta_{xy}$, may break $[C_2||M_{xy}]$ as well, leading to an effective ferromagnetic SPG that has no symmetry connecting opposite spin sublattices. Therefore, in this case, the spin polarized Fermi surface is similar to that for an FM, as shown in Fig.~\ref{p_s}(f). Table II lists the effective SPGs of $g$- and $i$-wave AMs under linear strain ($\eta_{ij}$) by collecting spin group symmetry operations $[R_1||R_2]$ such that $R_2\eta_{ij}\rightarrow \eta_{ij}$.
 
Our symmetry analysis shows how external stimuli, which break spin group symmetries that enforce zero spin conductivity, can be used to allow spin conductivity in a larger group of materials. It should be further noted that the symmetry-lowering stimuli in Tables I and Table II can be applied simultaneously to obtain spin conductivities in conventional magnets through piezomagnetoelectric-like ($\mathcal{E}_i\eta_{jk}$) couplings, where electric and strain fields act in concert. For example, AFM SPG $^14/^2m ^2m ^2m$ allows spin conductivity under the application of $\mathcal{E}_z\eta_{ij}$, but not when $\mathcal{E}_z$ or $\eta_{ij}$ are used separately.


\section{DFT Analysis}
In this section, we use DFT-based calculations to investigate spin current generation in selected materials that are representative of the distinct symmetry cases summarized in Tables I and II. We begin by analyzing the spin conductivity in both bulk and two-dimensional $d$-wave AMs, which intrinsically support finite spin currents. Subsequently, we consider prototypical materials to explore three key mechanisms for spin current generation: (i) an electric field–driven transition from antiferromagnetic (AFM) to uncompensated magnetic order, (ii) a strain-induced symmetry transition from $g$-wave to $d$-wave AM, and (iii) a combined electric field and strain-mediated transition from $i$-wave to $d$-wave altermagnetism. Note that in accordance to Neumann's principle, we only need to consider point group operations for the response tensors that represent macroscopic/bulk properties~\cite{cracknell1973symmetry}. Hence, we have omitted the translation part from the symmetry operations. Also, we have omitted the spin-only symmetry group ``$^{\infty m}1$" from our discussion of SPGs as this continuous symmetry only enforces conditions on the transverse spin polarization response ($\sigma^{x/y}_{ij}=0$), but imposes no conditions on $\sigma^z_{ij}$. Additionally, we define the charge-to-spin conversion ratio as $\frac{2e}{\hbar}\sigma^z_{ij} / \sigma_{ij}$~\cite{belashchenko2025giant}. In the limiting case, when energy falls into a region without bands, the charge conductivity $\sigma_{ij}$ approaches zero, and this ratio becomes mathematically indeterminate. To address this, we set the conversion ratio to zero whenever $\sigma_{ij}$ falls below a defined threshold in the charge-to-spin conversion versus energy plots. Note that in semiconductors, hole or electron doping is required to observe spin transport. A low-level doping concentration ($\approx 2$ \%) is sufficient to provide a stable carrier density while maintaining the compensated magnetic order~\cite{galindez2025revealing}.

\subsection{$d$-wave AMs}
\begin{figure}
	\includegraphics[width=3.5in]{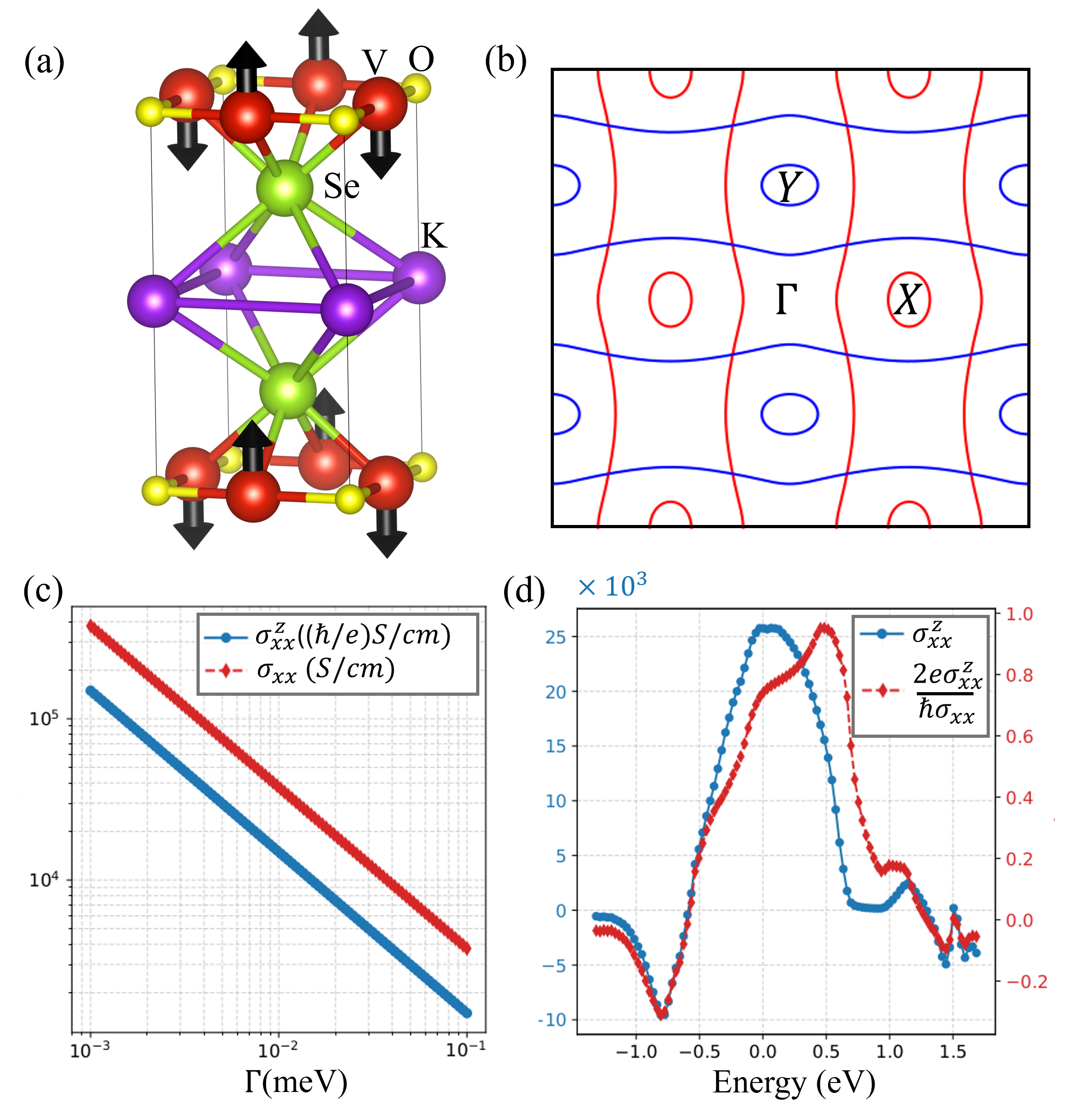}
	\caption{(a) The crystal and magnetic structures of KVSe$_2$O. (b) The Fermi surface of KVSe$_2$O projected on the $k_z=0$ plane, calculated without spin-orbit coupling. The red and blue curves denote spin-up and spin-down Fermi surfaces, respectively. (c) The charge and spin conductivity as a function of the scattering rate ($\Gamma$). (d) The spin conductivity and corresponding charge-to-spin conversion ratio as a function of the Fermi energy.}

    \label{p2}
\end{figure}
\begin{figure}[t]
	\includegraphics[width=3.5in]{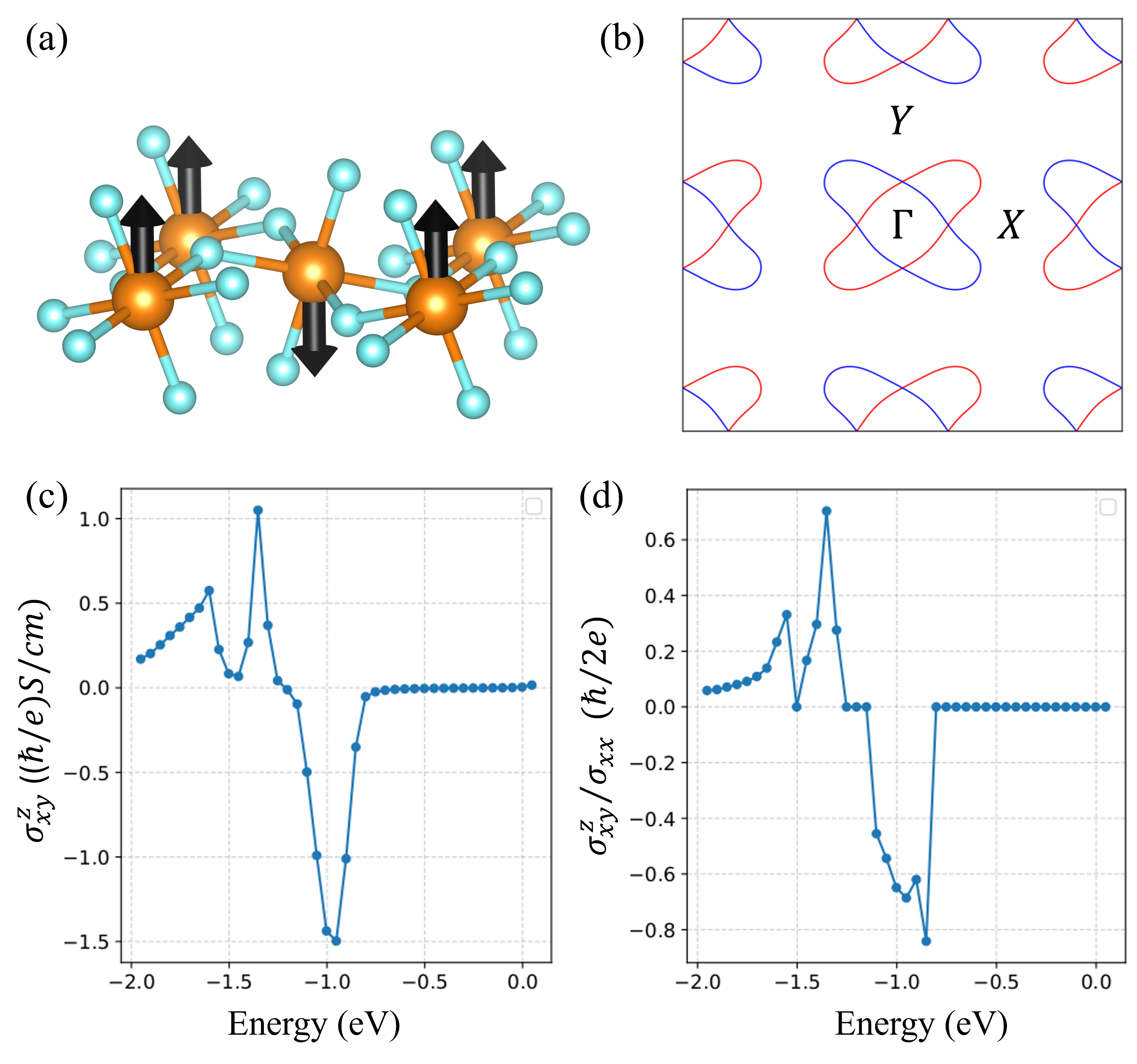}
	\caption{(a) The crystal and magnetic structures of RuF$_4$. (b) The Fermi surface of RuF$_4$ projected on the $k_z=0$ plane, calculated without spin-orbit coupling. (c) The spin conductivity and (d) corresponding charge-to-spin conversion ratio as a function of the Fermi energy.}
    \label{p3}
\end{figure}
KVSe$_2$O belongs to a broader class of materials crystallizing in the orthorhombic crystal structure with a space group of $P4/mmm$ [see Fig.~\ref{p2}(a)]. KVSe$_2$O is a room temperature metallic altermagnet with quasi-one-dimensional Fermi sheets extending along the $k_x$ and $k_y$ direction, and quasi-2D Fermi pockets centered around the X and Y points as seen in Fig.~\ref {p2}(b). The SPG of KVSe$_2$O is $^24/^1m^1m^2m$, containing $[C_2||C_{4z}]$, $[E||P]$, $[E||M_z]$, $[E||M_x]$, and $[C_2||M_{xy}]$. Following Neumann's principle, the symmetries of this material's SPG are adopted by its spin anisotropy. Using the transformation rules of Eq. 1 and Eq. 2, $[C_2||C_{4z}]$ transforms $\sigma^{s_z}_{yy}=\sigma^{-s_z}_{xx}$, $\sigma^{s_z}_{xy}=-\sigma^{-s_z}_{yx}$, $\sigma^{s_z}_{zx}=\sigma^{-s_z}_{zy}$, and $\sigma^{s_z}_{zy}=-\sigma^{-s_z}_{zx}$. Therefore, $[C_2||C_{4z}]$ enforces $\sigma^{z}_{xz}=\sigma^{z}_{yz}=\sigma^{z}_{zx}=\sigma^{z}_{zy}=\sigma^{z}_{zz}=0$. Furthermore, $[C_2||M_{xy}]$ transforms $\sigma^{s_z}_{xy} \rightarrow \sigma^{-s_z}_{yx}$, and hence, $\sigma^{z}_{xy}=\sigma^{z}_{xy}=0$ (Eq. 2). Following a similar procedure for all symmetry operations, the symmetry allowed components of the spin conductivity tensor are  $\sigma^z_{yy}$ and $\sigma^z_{xx}$, with $\sigma^z_{yy}=-\sigma^z_{xx}$. Figure~\ref{p2}(c) shows the spin conductivity, $\sigma^z_{xx}$, for the different scattering rates. The spin conductivity is inversely proportional to $\Gamma$ within the considered range of 1 meV to 100 meV. The charge-to-spin conversion efficiency ($\frac{2e}{\hbar}$ $\sigma^k_{ij}/\sigma_{ij}$) is 76\%, which can reach almost 100\% with electron doping of 0.5 eV [see Fig.~\ref{p2}(d)], where the Fermi pockets around the X and Y points vanish and spin currents are entirely governed by the Fermi sheets .

RuF$_4$ is a well known 2D $d$-wave AM crystallizing in space group $P2_1/c$ [see Fig.~\ref{p3} (a)]. The SPG of RuF$_4$ is $^22/^2m$ with spin symmetry operations $[C_2||M_y]$ and $[E||P]$ [see Fig.~\ref{p3} (b)]. $[C_2||M_y]$ transforms $\sigma^{s_z}_{ii}=\sigma^{-s_z}_{ii}$, therefore $\sigma^{z}_{ii}=0$. Also, $[C_2||M_y]$ leads to  $\sigma^{s_z}_{xy}=-\sigma^{-s_z}_{xy}$, $\sigma^{s_z}_{yx}=-\sigma^{-s_z}_{yx}$, $\sigma^{s_z}_{xz}=\sigma^{-s_z}_{xz}$,
$\sigma^{s_z}_{zx}=\sigma^{-s_z}_{zx}$, $\sigma^{s_z}_{yz}=-\sigma^{-s_z}_{yz}$, and
$\sigma^{s_z}_{zy}=-\sigma^{-s_z}_{zy}$. Therefore, the nonzero components of the spin conductivity tensor are $\sigma^{z}_{xy}$, $\sigma^{z}_{yx}$, $\sigma^{z}_{yz}$, and $\sigma^{z}_{zy}$. 
Since RuF$_4$ is a semiconductor, the charge conductivity is zero at the Fermi energy.  Fig. 3(c) shows the spin conductivity, $\sigma^z_{xy}$, of the hole-doped RuF$_4$ as a function of the Fermi energy. The maximum charge-to-spin conversion efficiency is 80\% when the hole doping moves the Fermi level below the intrinsic Fermi level by $\sim$0.9 eV [Fig. 3(c)].

Note that although both KVSe$_2$O and RuF$_4$ are $d$-wave AMs, their respective responses to the applied electric field differ from each other. This is owing to the different spin anisotropies of their Fermi surfaces. In the case of RuF$_4$, the bands are spin-degenerate along the $x$ and $y$ directions, while for KVSe$_2$O, the bands are spin-degenerate along the $x+y$ and $x-y$ directions. Hence, if the applied electric field is in the $x$ direction, KVSe$_2$O will show longitudinal spin conductivity while RuF$_4$ will exhibit transverse spin conductivity. On the other hand, KVSe$_2$O will exhibit transverse spin conductivity, and RuF$_4$ will show longitudinal spin conductivity when an electric field is applied along the $x+y$ or $x-y$ direction. 
The longitudinal and transverse spin conductivity for an arbitrary direction of the electric field for KVSe$_2$O and RuF$_4$ is plotted in Section I of the SM.

\begin{figure}
	\includegraphics[width=3.3in]{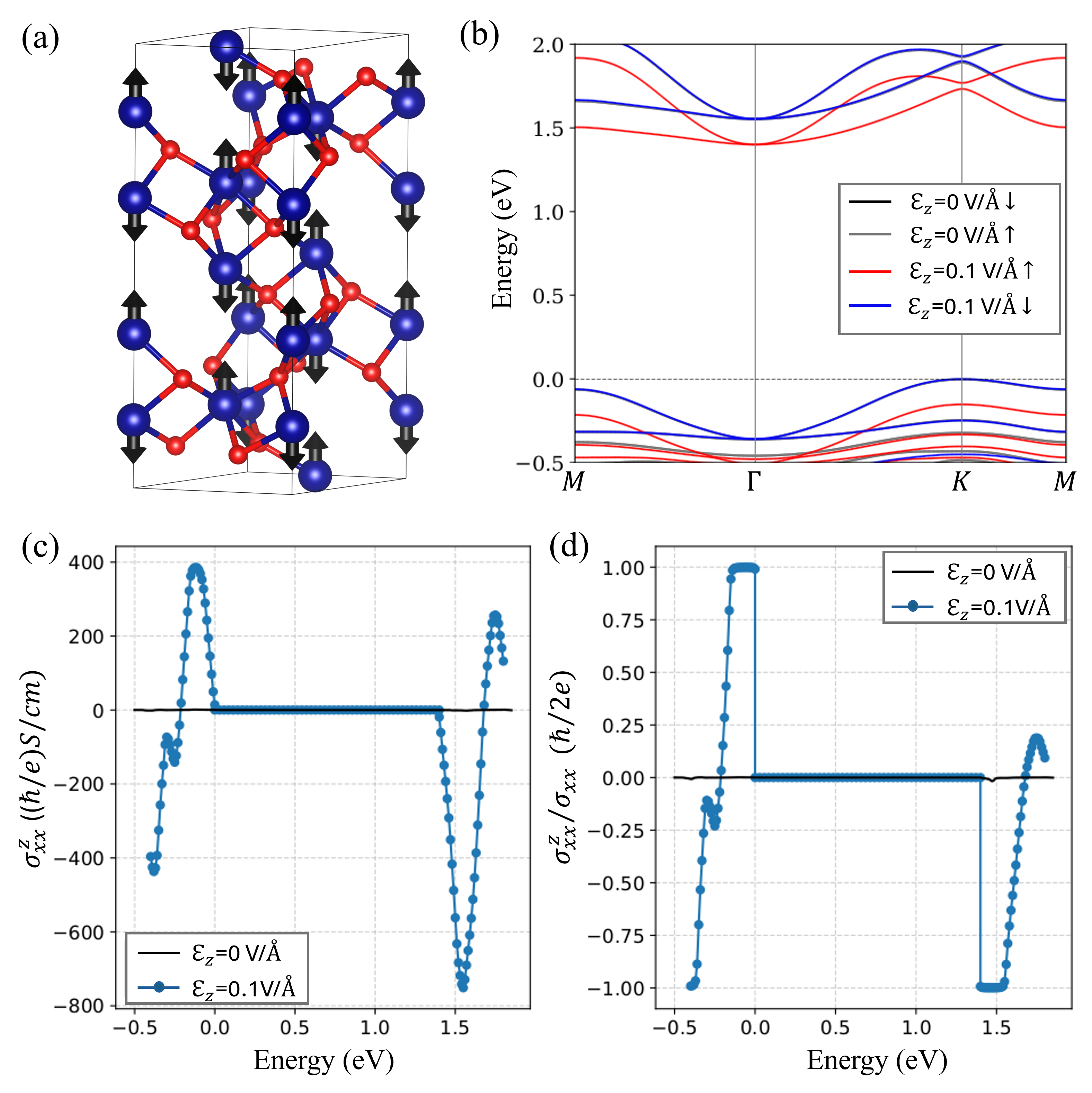}
	\caption{(a) The crystal and magnetic structures of Cr$_2$O$_3$. The blue and red spheres are used for Cr and O atoms, respectively. (b) Spin-polarized band structure of [0001] film of Cr$_2$O$_3$ with and without the electric field, $\mathcal{E}_z$. (c) The spin conductivity and (d) corresponding charge-to-spin conversion ratio as a function of the Fermi energy with and without $\mathcal{E}_z$.}
    \label{p4}
\end{figure}
\subsection{$\mathcal{E}_z$ induced AFM to uncompensated magnetism}
Bulk Cr$_2$O$_3$ is a well-known magnetoelectric antiferromagnet that crystallizes in the $R\overline{3}c$ space group symmetry. The SPG of Cr$_2$O$_3$ is $^2\overline{3}{^1}1^1m$ with the following spin symmetry operations: $[C_2||P]$, $[E||M_y]$, $[E||C_{3z}]$, and $[C_2||S_{3z}]$ [see Fig.~\ref{p4}(a)]. Since the $[C_2||P]$ symmetry operation transform $\sigma^{s_z}_{ij}$ to  $\sigma^{-s_z}_{ij}$, spin conductivity is forbidden in bulk Cr$_2$O$_3$. 
The application of $\mathcal{E}_z$ breaks $[C_2||S_{3z}]$ and $[C_2||P]$, while preserving $[E||M_y]$ and $[E||C_{3z}]$ symmetries. This reduces the original SPG to a ferromagnetic  $^13^11^1m$ SPG (see Table I). To apply the electric field, we used a [0001] film of Cr$_2$O$_3$ with a 20\,\AA{} vacuum separating the film from its image in the $z$-direction. This structure was previously found to result in the most stable surface~\cite{wysocki2012microscopic}.  Since the electric field, $\mathcal{E}_z$, breaks the $[C_2||P]$ symmetry, which is the only symmetry that connects the opposite spin sublattices, it leads to a Zeeman-like spin splitting of the conduction and valence bands, as seen in Fig. 4(b). Despite the presence of spin splitting, which is similar to that of a ferromagnet, our DFT calculations reveal a net zero magnetic moment. This apparent contradiction can be attributed to the semiconducting nature of Cr$_2$O$_3$, which gives rise to zero net magnetic moment according to the Luttinger's criterion.  In a stoichiometric compound, Luttinger's criterion requires that the net magnetization be equal to $N\,\mu_{B}$ ($N$=integer). When at least one spin channel is gapped (i.e. the material is either a semiconductor or a half metal), and the orbital moment is quenched, this results in zero net magnetic moment for small perturbations.  The $[E||M_y]$ and $[E||C_{3z}]$ symmetries allow spin conductivity components -- $\sigma^{z}_{xx}$, $\sigma^{z}_{yy}$, and $\sigma^{z}_{zz}$ -- to be non-zero, with $\sigma^{z}_{xx}=\sigma^{z}_{yy}$. It should be pointed out that since $J^z_x=\sigma^z_{xx}\mathcal{E}_x$ and $\sigma^z_{xx}$ itself depends on $\mathcal{E}_z$, we get that $J_x \propto \mathcal{E}_x \mathcal{E}_z$. In other words, the spin conductivity is at least second order in the electric field.  Figure 4(c) plots the spin conductivity, $\sigma^z_{xx}$, for the electron- and hole-doped Cr$_2$O$_3$ as a function of Fermi energy with and without an external electric field ($\mathcal{E}_z$ = 0.1 V/{\AA}).  The spin conductivity $\sigma^z_{xx}$ switches signs on going from negative values of Fermi energy (corresponding to hole-doping) to the positive values (electron doping). This is owing to the opposite spin magnetization of the topmost valence and lowest conduction bands. As seen in Fig. 4(c), the maximum value of $\sigma^z_{xx}$ is $\sim 750(\hbar/e) S/cm$ for the electron-doped material. Fig. 4(d) plots the charge-to-spin conversion ratio as a function of Fermi energy. It is clearly seen that the charge current is fully spin polarized around the CBM and VBM, as only one type of spin-conduction channel contributes to the conductivity. On the other hand, away from the CBM and VBM, the charge-to-spin conversion ratio decreases as both spin conduction channels start contributing.

\begin{figure}
	\includegraphics[width=3.5in]{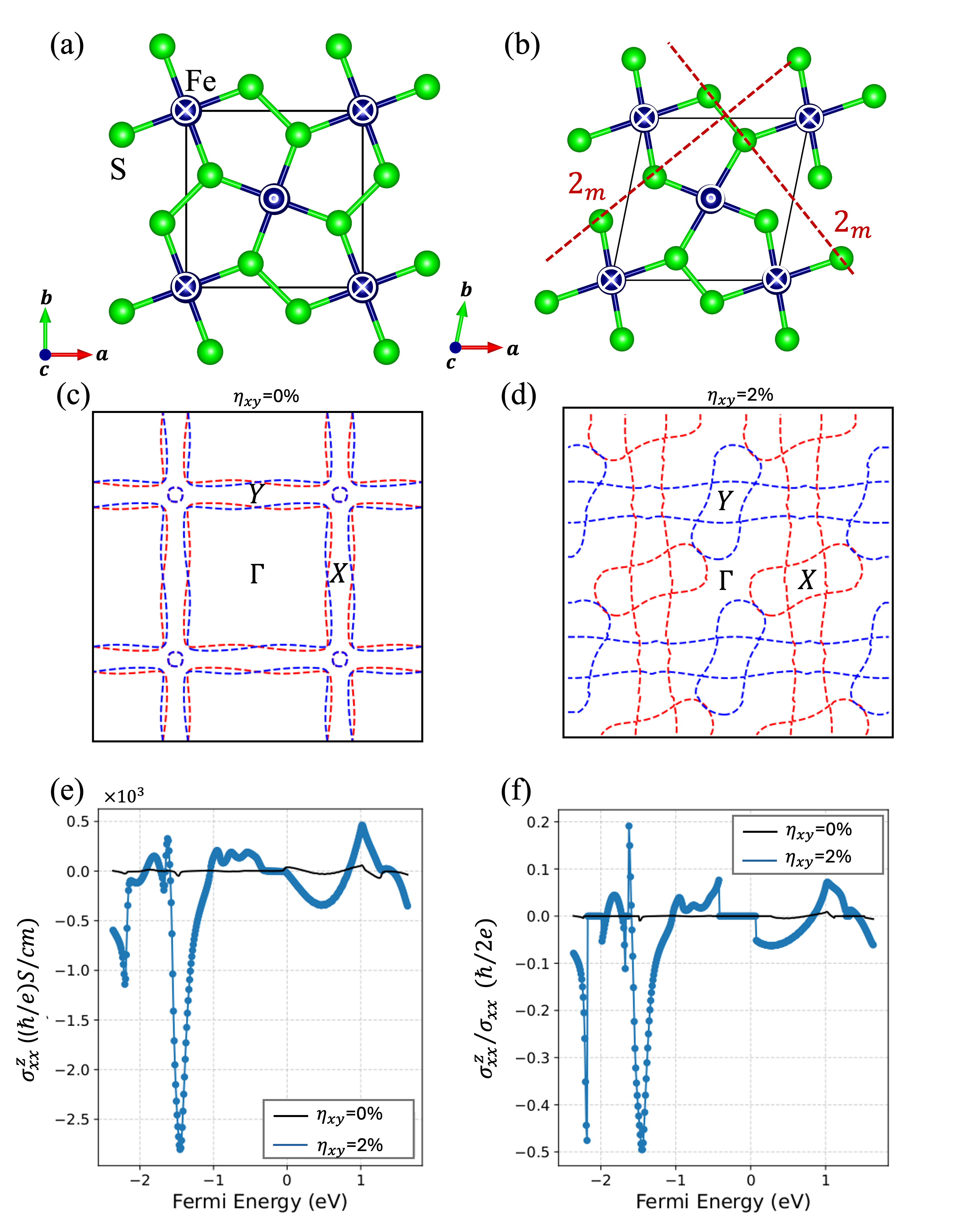}
\caption{(a) The crystal and magnetic structures of an FeS$_2$ monolayer (a) without and (b) with strain field, $\eta_{xy}$. The magnetic structures in (a) and (b) are shown by superimposing the $\odot$ and $\otimes$ symbols on the Fe atoms, representing spins that point in $+z$ (i.e out of the plane of paper) and $-z$ (into the plane of paper) directions, respectively. Also, in order to make the strain-induced distortions visible to the naked eye, we have exaggerated the strain to 20\% in (b). The constant energy contours for (c) unstrained and (d) strained FeS$_2$ for E$_F-1.45$\,eV with maximum charge-to-spin conversion. The (e) spin conductivity and (f) charge-to-spin conversion ratio as a function of energy.}
    \label{p5}
\end{figure}

\begin{figure}
	\includegraphics[width=3.5in]{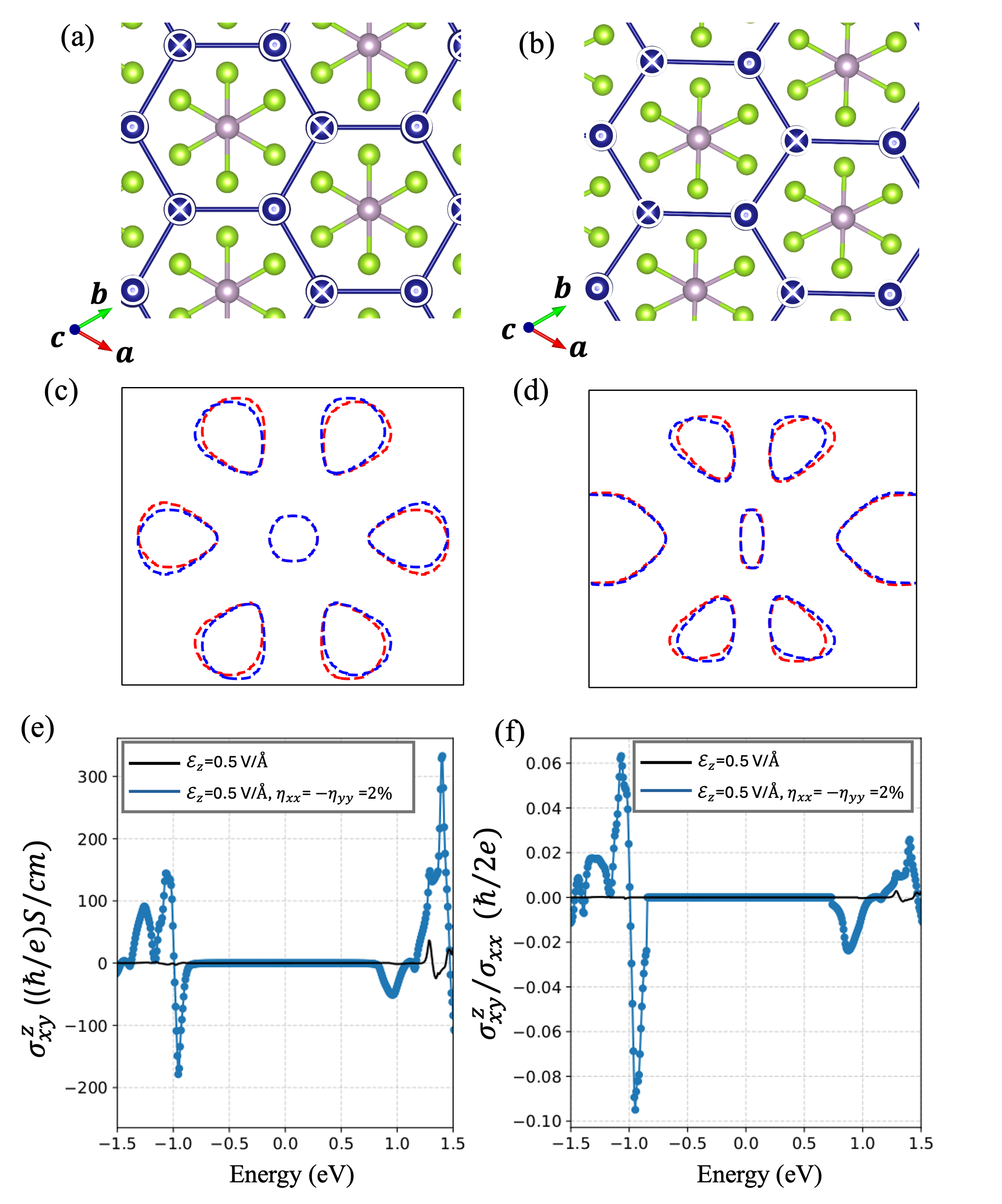}
\caption{(a) The crystal and magnetic structures of the monolayer MnPSe$_3$ (a) without and (b) with applied strain.  In (b), the strain, $\eta_{xx}+\eta_{yy}$ (with $\eta_{xx}=-\eta_{yy}$), is exaggerated to 10\% to magnify the structural distortions, making them visible to the naked eye.  Also, the magnetization in (a) and (b) are shown by superimposing the $\odot$ and $\otimes$ symbols on the Mn atoms, representing spins that point in $+z$ (i.e out of the plane of paper) and $-z$ (into the plane of paper) directions, respectively. The constant energy contours (at an energy of E$_\textrm{F}-1.02$ eV) for (c) $\mathcal{E}_z=0.5$ V/{\AA} and (d) $\mathcal{E}_z=0.5$ V/{\AA}, and $\eta_{xx} = -\eta_{yy}$ = 0.02, showing the $i$ to $d$ wave transition. The (e) spin conductivity and (f) charge-to-spin conversion ratio as a function of energy.}
    \label{p6}
\end{figure}

\subsection{$\eta_{xy}$ induced $g$-wave to $d$-wave transition}

FeS$_2$ {belongs} to a large class of planar 2D materials that crystallize in the $P4/mbm$ space group [Fig. 5(a)]. FeS$_2$ is a planar $g$-wave AM with nontrivial SPG $^14/^1m^2m^2m$ containing $[C_2||M_x]$, $[C_2||M_{xy}]$, $[E||C_{4z}]$, and $[E||M_z]$. As discussed previously, the symmetry operation $[C_2 || M_x]$ suppresses the {following spin conductivity components}: $\sigma^z_{xx}$, $\sigma^z_{yy}$, $\sigma^z_{zz}$, $\sigma^z_{yz}$, and $\sigma^z_{zy}$. Additionally, the mirror symmetry, $[E || M_z]$, eliminates $\sigma^z_{xz}$ and $\sigma^z_{zx}$, while $[C_2 || M_{xy}]$ forbids $\sigma^z_{xy}$ and $\sigma^z_{yx}$. Therefore, spin current is completely forbidden in a planar $g$-wave AM. Consistent with the symmetry analysis, DFT calculations confirm a vanishing spin conductivity {[see the solid black line in Fig. 5(e)]. The $g$-wave SPG $^14/^1m^2m^2m$ of FeS$_2$ reduces to a $d$-wave SPG ($^2m^2m^1m$) through application of $\eta_{xx}$, $\eta_{yy}$, or $\eta_{xy}$ strains, whereas it can be reduced to a $d$-wave SPG ($^22/^2m$) through an application of $\eta_{xz}$ or $\eta_{yz}$ strains [see Table II].  As a representative case, we use the shear strain, $\eta_{xy}$, of 2\%. The structural changes produced by this strain can be seen in Fig. 5 (b), where we have exaggerated the strain percentage to 20\% for the sake of representation only. The SPG of FeS$_2$ under $\eta_{xy}$ strain is $^2m^2m^1m$, with spin symmetry operations $[C_2||M_{xy}]$, $[C_2||M_{xy'}]$, and $[E||M_{z}]$.} Therefore, $^2m^2m^1m$ allows for non-zero $\sigma^z_{xx}$ and $\sigma^z_{yy}$ components, with $\sigma^z_{yy}=-\sigma^z_{xx}$.  Figures 5(c) and 5(d) show the altermagnetic phase transition from planar $g$-wave to planar $d$-wave upon the application of the strain. The maximum spin conductivity value is calculated to be upto 2800\,$\frac{\hbar}{e}\frac{S}{cm}$ as seen in 5(e). This occurs at an energy of -1.45\,eV, where the Fermi surface is mostly composed of spin-polarized sheets along the $x$ and $y$ directions [Figs. 5(d) and 5(e)]. These Fermi sheets are similar to those for the case of KVSe$_2$O [Fig. 2(b)]. As seen in Fig. 5(f), the maximum spin-to-charge conversion ratio is 50\%, which is smaller than that obtained for KVSe$_2$O due to the Fermi pockets around $X$ and $Y$, which are slighter bigger than those of KVSe$_2$O [compare Figs. 2(b) and 5(d)]. Recall from subsection A that transverse spin conductivity will be observed when the electric field is applied along the $x\pm y$ directions [see section I of SM]. 

\subsection{ $\mathcal{E}_z\eta_{xx}$ induced AFM to $d$-wave AM transition}
MnPSe$_3$ is a 2D antiferromagnet, which crystallizes in the $P\overline{3}1m$ space group [Fig.~\ref{p6}(a)]. The SPG of MnPSe$_3$ is $^2\overline{3}{^2}m{^1}1$, and spin current is forbidden due to spin-degenerate bands. MnPSe$_3$ has been the subject of several reports that have studied its AFM to $d$-wave AM transition under the application of a vertical electric field~\cite{sheoran2025spontaneous, sheoran2024nonrelativistic}. Under the application of an electric field ($\mathcal{E}_z$), the SPG of MnPSe$_3$ reduces from $^2\overline{3}{^2}m{^1}1$ to $^1{3}^2m{^1}1$, with the latter containing $[E||C_{3z}]$ and $[C_2||M_x]$ symmetries [see Table I and Fig.~\ref{p6}(b)]. Although the $^1{3}{^1}1^2m$ SPG corresponds to the bulk $g$-wave AM, the 2D nature of MnPSe$_3$ ($k_z=0$) makes it a planar $i$-wave AM [Fig.~\ref{p6}(c)]. The $i$-wave altermagnetic nature of monolayer MnPSe$_3$ can be confirmed from the six-fold symmetric constant energy-contours in Fig.~\ref{p6}(c).  The spin currents are forbidden with and without the electric field due to underlying spin group symmetries $[E||C_{3z}]$ and $[C_2||M_x]$. This is due to the fact that $[E||C_{3z}]$ enforces $\sigma^{s_z}_{xx}=\sigma^{s_z}_{yy}$, $\sigma^{s_z}_{xy}=-\sigma^{s_z}_{yx}$, and $\sigma^{s_z}_{xz}=\sigma^{s_z}_{yz}=\sigma^{s_z}_{zx}=\sigma^{s_z}_{zy}=0$, while the $[C_2||M_x]$ symmetry leads to $\sigma^{s_z}_{ii}=\sigma^{-s_z}_{ii}$, $\sigma^{s_z}_{yz}=\sigma^{-s_z}_{yz}$, $\sigma^{s_z}_{zy}=\sigma^{-s_z}_{zy}$, $\sigma^{s_z}_{xy}=-\sigma^{-s_z}_{xy}$, $\sigma^{s_z}_{yx}=-\sigma^{-s_z}_{yx}$, $\sigma^{s_z}_{xz}=-\sigma^{-s_z}_{xz}$, and $\sigma^{s_z}_{zx}=-\sigma^{-s_z}_{zx}$.  However, the $^1{3}^2m{^1}1$ SPG can be further reduced to the  $^2m$ SPG by using $\eta_{xx}$, $\eta_{yy}$, and $\eta_{yz}$ [Table II]. On the other hand, an application of either $\eta_{xy}$ or $\eta_{xz}$  reduces the SPG to that of an uncompensated magnet, while the $^1{3}^2m{^1}1$  SPG remains unaltered by the uniaxial strain along the $z$-direction ($\eta_{zz}$). Therefore, monolayer MnPSe$_3$ AFM will undergo a magnetic transition to a $d$-wave altermagnet under $\mathcal{E}_z\eta_{xx}$, $\mathcal{E}_z\eta_{yy}$, and $\mathcal{E}_z\eta_{yz}$. For illustrative purposes, we consider $\mathcal{E}_z(\eta_{xx}+\eta_{yy})$ with $\eta_{xx}=-\eta_{yy}= 0.02$, resulting in the structure shown in Fig.~\ref{p6}(b), and the corresponding constant energy contour in Fig.~\ref{p6}(d). The new SPG of MnPSe$_3$ under this strain is $^2m$, containing the $[C_2||M_x]$ symmetry operation [see Tables I and II]. As mentioned earlier, the only nonzero components of the spin conductivity tensor are $\sigma^{z}_{xy}$, $\sigma^{z}_{yx}$, $\sigma^{z}_{yz}$, and $\sigma^{z}_{zy}$. The  $\sigma^{z}_{yz}$ and  $\sigma^{z}_{zy}$ components are forbidden due to the 2D nature of MnPSe$_3$. Figures ~\ref{p6}(e) and ~\ref{p6}(f) show the spin conductivity and charge-to-spin conversion ratio for MnPSe$_3$ with electric field $\mathcal{E}_z$ and $\mathcal{E}_z(\eta_{xx}+\eta_{yy})$. While there is no spin current in the absence of strain, the strain-induced charge-to-spin conversion ratio reaches up to $\sim$10\%, which  corresponds to a spin-splitter angle of approximately $11^{\circ}$. Here we have used the following expression for the spin-splitter angle: \(
\theta_{ij} = \arctan\!\left( 2\,\frac{2e}{\hbar}\,\frac{\sigma^{s}_{ij}}{\sigma^{c}_{ii}} \right)
\)~\cite{gonzalez2021efficient,karetta2025strain}, with $\sigma^{s}_{ij}$ ($\sigma^{c}_{ij}$ ) being the transverse spin conductivity 
(longitudinal charge conductivity). This spin-splitter angle is experimentally measurable~\cite{karetta2025strain, tao2018self}.

\section{Conclusion}

In this work, we have proposed a general framework to obtain nonrelativistic spin currents in collinear AFMs, and $g$- and $i$-wave AMs. Using symmetry analysis, we have demonstrated that spin conductivity can be induced through magnetoelectric, piezomagnetic, and piezomagnetoelectric-like couplings.  The exact functional form of the spin-conductivity tensor is determined by the effective reduced SPG under the application of an external electric field and/or strain.  By performing DFT calculations on realistic materials that were subjected to either electric fields, strains, or both, we have shown that the charge-to-spin conversion ratios in AFMs, as well as $g$- and $i$-wave AMs, can be as high as those seen for ferromagnets, compensated ferrimagnets, and $d$-wave altermagnets.  It should be pointed out that the external stimuli considered in our work are realistic and can be achieved experimentally.  For example, the direction-dependent strain field can be obtained by transferring the exfoliating altermagnetic crystals onto a flexible substrate with pre-patterned metal pads~\cite{son2019strain,sheoran2023probing}.  The feasibility of our proposed approaches is also confirmed by recent reports of in-plane strain and magnetic field-induced nematicity in AFM intercalated transition metal dichalcogenides~\cite{feng2025nonvolatile}. In addition, recently, finite anomalous Hall conductivity was successfully measured in the semiconducting $g$-wave AM, $\alpha$-Fe$_2$O$_3$, with 1\% Ti doping~\cite{galindez2025revealing}. Hence, doping to induce charge conductivity, along with electric fields and/or strain to drive the symmetry-lowering phase transitions, can be used to observe spin conductivity in intrinsically-semiconducting AFMs and higher-order AMs. Our proposed design principles for inducing spin currents are general and will expand the range of materials available for ultrafast magnetic spintronics.

\section*{Acknowledgments}  
The research performed at Howard University was supported by Air Force Office of Scientific Research under award number FA9550-23-1-0679 and the National Science Foundation Grant No. DMR-1752840. This work used the Expanse cluster at SDSC through allocation PHY180014 from the Advanced Cyberinfrastructure Coordination Ecosystem: Services \& Support (ACCESS) program, which is supported by National Science Foundation Grants Nos. 2138259, 2138286, 2138307, 2137603, and 2138296, and Maryland Advanced Research Computing Center. 


\bibliography{ref}

\end{document}